# Possibilities for a Causal Interpretation of Wave Mechanics

D.H. Delphenich[†]
Physics Department
University of Wisconsin – River Falls
River Falls, WI 54022

*Abstract. The basic physical problems that necessitated the emergence of quantum physics are summarized, along with the elements of wave mechanics and its traditional statistical interpretation. Alternative interpretations to the statistical one, such as the hydrodynamical and optical interpretations, nonlinear waves and nonlinear electrodynamics, and the conception of spacetime as an ordered medium are reviewed.*

**0. Introduction.** Many years have passed since the the Solvay Conference of 1927, when the Copenhagen School first enshrined the statistical interpretation of wave mechanics as the reigning dogma of quantum physics. Nevertheless, there are many reasons for continuing to maintain that the most fundamental problems of physics that early quantum theory addressed remain just as unresolved in the modern era as they were in the 1920's. The main reasons for the widespread acceptance of that statistical interpretation, along with its attendant philosophy of phenomenology, seem to have been the usual combination of social inertia and the continued absence of any competing interpretation that would present a more fundamental character in a convincing way.

It is important to pose those fundamental problems of quantum physics in a manner that ignores the time that has passed since the early years of quantum theory, in order to emphasize the enigmatic character of the problems and the fact that the mathematical models of physics are ultimately just an endless sequence of concentric spheres that each enclose all of the preceding theoretical models, but always at the expense of simplifying approximations that inevitably define the limits of applicability of the model to the phenomena of nature.

We state the basic problems of quantum physics in such a way as to emphasize the fact that what they all have in common is that they are all problems in the theory of the electromagnetic interaction:
 *a)* The structure of the source charge distributions and fields of static elementary charges.
 *b)* The field of a moving elementary charge, especially an accelerated one.
 *c)* The structure of the bound states of systems of elementary charges.

The first problem was generally treated under the heading of "the theory of the electron," since most of its definitive efforts preceded the discovery of other charged particles besides electrons and protons; in particular, the discovery of antiparticles was still far in the future. The main contributors to that theory were Abraham, Lorentz, and Poincaré [**1-3**]. The key issue that they addressed was the tradeoff between considering elementary charges as having a pointlike charge distribution or having a spatially extended one. Subordinate to that decision were the problems of possibly attributing the mass of the charge to the total energy in the field that it produced and ensuring the stability of the charge distribution. Although a pointlike charge distribution had the desired stability, it led to an infinite self-mass for the electron. On the other hand, the extend charge models generally satisfied the requirement of a finite self-mass, but at the expense of stability, at least in the absence of further assumptions about the nature of the binding force that stabilized the distribution against its mutual Coulomb repulsion. Note that precisely the same question arises in the context of the stability of nuclei under the mutual repulsion of their constituent protons. It is ironic that although one could say that quantum chromodynamics resolved the issue of what sort of force stabilizes nuclear charge distributions, nevertheless, the least understood problems of strong interaction physics, namely, low energy quantum chromodynamics, include precisely that problem.

---

[†] david.delphenich@uwrf.edu

The second and third problems are both concerned with the radiative field modes of elementary charges; in one case, the scattering states, and, in the other, the bound states. Although it was agreed that the motion of an accelerated charge should be accompanied by an emitted radiation field (i.e., a photon) - for instance, photons are produced by the oscillation of electrons in antennas and in the form of the brehmsstrahlung emitted by decelerating charges, nevertheless, the precise manner by which this happened was incompletely understood. Dirac [**3,4**] proposed an extension to the Lorentz force law that included the radiation reaction associated with the acceleration of a charge, but the fact that it also included the third derivative of position, and thus raised the order of the dynamical equation above the customary two, led to various unacceptable pathological solutions. For instance, one had to contend with runaway electrons and pre-acceleration under collision. Also, there was the question of whether a constant acceleration would produce radiation.

The bound state problem was the one that defined the historical course of quantum physics most definitively, because it include the problem of the spectrum of blackbody radiation and the spectra of atomic electrons. In both cases, one had to contend with discreteness in their nature, which was completely beyond the scope of Maxwell's theory of electromagnetism.

This last statement brings us to the essence of the fundamental problems of quantum physics: they are all associated with the limits of Maxwell's theory. For instance, some of the limiting assumptions that it implicitly contained were: the idea that the constitutive law of the vacuum could be summarized into two constants – i.e., the vacuum dielectric constant $\varepsilon_0$ and magnetic permeability $\mu_0$, the consequent linearity in the vacuum form of its equations, and the impossibility of indefinitely extending the empirically obtained form of the Coulomb field into the realm of small neighborhoods of elementary charges.

In all fairness to the memory of some of the giants of physics, the experimental discoveries that uncovered these limitations predated many of the advances of both physics and mathematics that are now associated with the aforementioned fundamental problems. For instance, in physics, one has the existence of antimatter and electron spin to account for, as well as the scattering of light by light and the fields of external potentials, not to mention the reducibility of the electronic charge into quark charges for hadronic matter. Also, there is an ongoing shift in the way that physics regards the structure of ordered media, which entails the appearance of topological defects and spontaneous symmetry breaking under phase transitions. In mathematics, much more is now known about geometry, topological, and nonlinear analysis than was available to the physicists of the Copenhagen School.

For completeness, we briefly summarize wave mechanics and its statistical interpretation:

The turning point in quantum theory seems to have been de Broglie's suggestion that just as Einstein had showed that the photon had particle-like properties, conversely, massive particles were also associated with a wavelike nature. The equation for the non-relativistic time evolution of these waves was obtained – by a largely heuristic process – by Schrödinger in the form:

$$\frac{h}{i}\frac{\partial \Psi}{\partial t} = -\frac{h^2}{2m}\Delta\Psi + U, \qquad (0.1)$$

in which $\Psi$ is the complex-valued wavefunction that describes a particle of mass $m$, and $U$ describes the force potential that acts upon it. This was extended to a relativistic form that described particles with integer spin by Klein and Gordon:

$$\Box\Psi + k_C^2\Psi = 0 \qquad (0.2)$$

in which $k_C = m_0 c/h$ is the Compton wavenumber for a particle of rest mass $m_0$, and ultimately (0.1) was given a relativistic form for half-odd-integer particles by Dirac:

$$\gamma^\mu \partial_\mu \psi + k_C \psi = 0, \qquad (0.3)$$

in which the $\gamma^\mu$ are the Dirac matrices that define the generators of a representation of the Clifford algebra of Minkowski space in the algebra of $4\times 4$ complex matrices, and $\psi$ is a spinor wavefunction that takes its values in $\mathbb{C}^4$, this time.

The main problem with Schrödinger's equation was finding some fundamental physical basis for explaining its empirical success in describing quantum phenomena. The Copenhagen School − mainly Bohr, Born, and Heisenberg – proposed the statistical interpretation, which gives the complex-valued quantum wavefunction $\Psi$ a sort of "pre-physical" character, like the potential 1-form $A$ of the electromagnetic field $F$, and resolved this by saying that all of the physical meaning is in the modulus-squared $|\Psi|^2$, which represents the probability density function for the position of the particle. The fact that $\Psi$ is complex is then attributed to the association of a phase factor with the wave/particle, as well.

Neither Einstein, nor even Schrödinger himself, ever completely accepted this statistical interpretation. Einstein saw its reliance upon the methods of probability and statistics as being a sign of incompleteness in the theory. Schrödinger never gave up hope that physics would find a more fundamental representation for his wavefunctions and the success of his equation.

No synopsis of the fundamental problems of quantum theory and electromagnetism would be complete without some mention of the fact that currently the reigning theory of electromagnetism at the atomic to sub-atomic level is quantum electrodynamics (QED). It is often incorrectly assumed that QED is a "complete" theory of electromagnetism, simply because it offers so many digits of accuracy in its agreement with the available experiments, and the fact that physics has yet to find an experiment that contradicts its predictions. The reason that this assumption is incorrect is because there are many problems that QED cannot address, due to its phenomenological "black-box" character. What it *can* do is provide an infinite sequence of perturbative refinements to the prediction of scattering data for elementary collision processes that involve charged particles and photons, along with somewhat dubious corrections to the inconsistencies that arise in the process of computing them. What it *does not* attempt to do is pose − much less solve − boundary value problems in the field equations that would describe these fundamental processes.

Objectively, though, QED also establishes certain aspects of electromagnetism as unavoidable in any theory that might presume to supersede it, such as the aforementioned matter/antimatter mirror symmetry in the elementary charges, and the closely associated process of vacuum polarization, and the idea that the ground state of a quantum harmonic oscillator − hence, a quantum electromagnetic field − is non-zero. The essence of these facts seems to suggest that the Maxwellian conception of the electromagnetic vacuum state was an over-simplification.

We shall now examine some of the historical attempts to avoid the statistical interpretation of wave mechanics, as well as some more modern observations and speculations that the author has been developing. The historical attempts we shall consider are the optical interpretations, the hydrodynamical intepretation, and the various attempts at nonlinear wave mechanics and nonlinear electrodynamics. Finally, we shall try to at least set the stage for the modeling of spacetime as an ordered medium that is capable of exhibiting phase transitions, as well.

**1. Optical interpretations.** The early hope of de Broglie [5] and Schrödinger [6] was that they could construct an exact analogy between the relationship of wave mechanics to particle mechanics and the relationship between wave optics and geometrical (ray) optics. The common element was the Hamilton-Jacobi equation [7]:

$$\begin{cases} \dfrac{\partial S}{\partial t} + H(t, x^i, p_i) = 0 \\ p_i = \dfrac{\partial S}{\partial x^i}. \end{cases} \quad (1.1)$$

From the standpoint of wave motion, the level surfaces of the Jacobi principal function $S = S(t, x^i)$ describe isophase hypersurfaces, and one can recover the particle trajectories from the characteristic equations of this first order partial differential equation, namely, Hamilton's equations:

$$\begin{cases} \dfrac{dx^i}{dt} = \dfrac{\partial H}{\partial p_i} \\ \dfrac{dp_i}{dt} = -\dfrac{\partial H}{\partial x^i} + \dfrac{\partial p_j}{\partial x^i}\dfrac{\partial H}{\partial p_j}. \end{cases} \quad (1.2)$$

By expressing these equations in latter form, we are, of course, anticipating the possibility that the matter whose momentum is described by $p_i$ is spatially extended, rather than pointlike.

The similarity between the first equation of (1.1) and Schrödinger's wave equation (0.1) is undeniable, except for the missing factor of $h/i$, which, of course, accounts for the difference between classical and quantum mechanics, as well. Ultimately, this analogy proved to be useful mostly in the geometrical optical, or WKB, approximation, in which one makes that limiting assumption that the wavelength of the wave is vanishingly small, i.e., $h \to 0$, in order that the wave phenomena associated with $S$, such as interference and diffraction, do not contradict the particle motion described by (1.2).

An interesting optical aspect of the Klein-Gordon equation (0.2) that defined one of the early attempts at giving a relativistic form to Schrödinger's equation is something that Klein himself observed [**8**]: equation (1.3) can be obtained from the five-dimensional linear wave equation by separating the fifth variable in the same way that Helmholtz's equation:

$$\Delta \Psi + k^2 \Psi = 0, \tag{1.3}$$

which is fundamental to the time-invariant formulation of wave optics, can be obtained from the four-dimensional wave equation. This carries with the problem of physically interpreting the fifth dimension that one has introduced. Since $k_C = h/m_0 c$ is the Compton wavelength associated with a particle of *total* rest mass $m_0$, we are suggesting that mass appears like a separation constant, i.e., an eigenvalue of something. One can make a good case for the notion that the fifth dimension that one has introduced is the proper time parameter, in such a way that the fifth component of the velocity becomes *c*. Another interesting aspect of this wave of representing matter waves is that although massive matter waves can propagate with any speed between 0 and *c*, non-inclusive, nevertheless, the five-dimensional wave that represents it must always propagate with the same characteristic velocity, like a massless particle. Further work on the physical details of five-optics was also carried out by Rumer [**9**], as well. Of course, Klein, along with Kaluza, Jordan, and Thirry examined the possibility that one could unify the theories of gravitation and electromagnetism by means of the geometry of five-dimensional Lorentzian manifolds (cf. [**10-11**]).

Later, we shall discuss possible analogies between nonlinear optics and QED in the context of nonlinear wave mechanics and nonlinear electrodynamics.

**2. Hydrodynamical interpretations for wave mechanics.** One of the most intriguing alternatives to the statistical interpretation to wave mechanics – i.e., the Schrödinger equation – was suggested in the same year that the Copenhagen School proposed their statistical interpretation. The alternative that Ernst Madelung [**12**] (see also Takabayasi [**13**]) suggested was a "hydrodynamical" interpretation that one obtained by first expressing the complex Schrödinger wavefunction in polar form $\Psi = e^{iS/h}$ and then separating the Scrödinger equation into real and imaginary parts. These equations take the form ([1]):

$$0 = \frac{\partial S}{\partial t} + \frac{1}{2m}(\nabla S)^2 - \frac{h}{2m}\frac{\Delta R}{R} \tag{2.1a}$$

$$0 = \frac{\partial R^2}{\partial t} + \frac{\partial}{\partial x^i}(\nabla S)^i. \tag{2.1b}$$

If one defines $\rho = mR^2$ to be the mass density function for a spatially extended object and uses the Hamilton-Jacobi prescription $p = dS = \partial_i S\, dx^i$ to define its momentum 1-form, and take the gradient of (2.1a) then these equations take the form:

$$0 = \frac{\partial p}{\partial t} + \frac{1}{m} p \cdot \nabla p + \nabla U_h \tag{2.2a}$$

$$0 = \frac{\partial \rho}{\partial t} + \frac{\partial p_i}{\partial x^i}. \tag{2.2b}$$

These equations seem to be describing an Euler-type equation (2.2a) for the motion of a continuous material medium of mass density $\rho$ and momentum $p$ under the action of a pressure term $\nabla U_h$, and a continuity equation (2.2b) for its mass. One can call this medium a "Madelung medium" or, more popularly, a "Madelung fluid," although Takabayasi pointed out that the stress tensor that one obtains for such a medium is not isotropic like an ideal fluid, nor does the stress couple to the rate of deformation like the viscosity in a viscous fluid. However, modern condensed matter physics has taken to using the term "hydrodynamics" in a broader sense than merely the dynamics of gases and liquids, so perhaps the term "Madelung fluid" is partially justified ([2]).

Some of the properties of this medium are: It has a flow velocity vector field defined by $\mathbf{v} = 1/\rho \nabla S$. It is irrotational in the sense that its dynamical vorticity 2-form $dp$ is zero, although if one forms the covelocity 1-form $U = 1/\rho\, dS$, one sees that its kinematical vorticity $du = -\rho^{-2} d\rho \wedge dS$ does *not* vanish. The steady flow states seem to correspond to the time-invariant solutions – i.e., eigenstates - to the Schrödinger equations.

When one applies the Madelung transformation to the Klein-Gordon equation, one obtains:

$$0 = p^2 + m_0^2 c^2 - h^2 \frac{\Box \sqrt{\rho_0}}{\sqrt{\rho_0}} \tag{2.3a}$$

$$0 = \delta p, \tag{2.3b}$$

in which the second equation also seems to define relativistic incompressibility as it was conceived by Lichnerowicz [**11**].

---

[1] We suppress any reference to the force potential $U$ insofar as it plays no essential role in the appearance of the quantum potential.

[2] It has been the observation of the author that modern physics - in particular, the theories of condensed matter, strings, and d-branes - seems to be quietly re-inventing the continuum mechanics that it so unjustly abandoned in the early Twentieth Century as a fundamental statement in physics.

So far, we have deferred mention of the "quantum potential" function that has appeared as a result of the Madelung process:

$$\begin{cases} U_h = -\dfrac{h^2}{2m}\dfrac{\Delta\sqrt{\rho}}{\sqrt{\rho}} & \text{non-relativistic case} \\ U_h^2 = -h^2 c^2 \dfrac{\Box\sqrt{\rho_0}}{\sqrt{\rho_0}} & \text{relativistic case.} \end{cases} \quad (2.4)$$

As the author showed in [**14**], there is a geometrical origin for this potential function, at least in the relativistic case: If one deforms the Minkowski space scalar product by a homothety $\eta \to g = \rho_0 \eta$ whose conformal factor $\rho_0$ is non-constant then the Levi-Civita connection for $g$ has a Ricci scalar curvature $\mathfrak{R}$ that satisfies:

$$U_h^2 = \tfrac{1}{6} h^2 c^2 \rho_0 \mathfrak{R}. \quad (2.5)$$

One might then regard the quantum potential energy as the work done in the process of deforming $\eta$ by way of the homothety. This seems to be consistent with Sakharov's conception of "metrical elasticity."

As was also shown in [**14**], one can further incorporate this deformation energy into the rest mass density of the medium by way of the definition:

$$\rho_0^2 = m_0^2 - \dfrac{1}{c^2} U_h^2, \quad (2.6)$$

which allows us to rewrite the transformed Klein-Gordon equation for the complex wavefunction $\Psi$ as equivalent to the system of equations in $\rho_0$ and $p$:

$$dp = 0, \quad \delta p = 0, \quad p^2 = \rho_0^2 c^2, \quad (2.7)$$

along with a consistency condition:

$$\Box R - \dfrac{m_0^2}{c^2} R(1-R^4) = 0, \quad (2.8)$$

in which we have reverted to the use of $R$ instead of $\rho_0$ for the sake of brevity. Notice that since $\rho_0 = R^2$, one can interpret $R$ as a dilatation that acts on tangent vectors (or frames) directly and produces the conformal transformation of the metric tensor field indirectly.

In the form (2.7), (2.8), the transformed equations clearly describe a mass distribution whose momentum propagates as a wave with non-characteristic speed and whose shape is dictated by a PDE of the nonlinear (real) Klein-Gordon type. Of course, one would ultimately have to account for the form of this latter equation in terms of deeper first principles, and not merely derive them from another foundation.

Another hydrodynamical interpretation of wave mechanics that attracted considerable attention was the manner by which Weyssenhoff [**15**] showed that the Dirac equation could be interpreted as equivalent to the equations of motion for a relativistic spinning fluid.

**3. Nonlinear wave mechanics and nonlinear electrodynamics.** One of the ways by which de Broglie envisoned extending wave mechanics was to a nonlinear theory [**16**], although one must really regard his work as more of a program than a theory. The essence of his *theory of the double solution* was that one might regard a wave/particle as the sum of a spatially extended wave function $\Psi = ae^{iS/\hbar}$ and a localized field $u = fe^{iS/\hbar}$ such that $\Psi$ obeyed the Klein-Gordon equation and the field $u$, which was assumed to embody the more singular nature of a localized particle, was to obey an unspecified nonlinear wave equation. The fact that we are assuming the consistency of the phases for both $\Psi$ and $u$ leads to the *guidance equation:*

$$\mathbf{v} = -\frac{1}{m}\nabla S, \qquad (3.1)$$

in which $\mathbf{v}$ is the velocity of $u$. Because the spatially extended wave seems to be guiding the particle along with its motion, $\Psi$ is sometimes referred to as the *pilot wave*.

One of the deep and far-reaching contributions of Jean-Pierre Vigier, whom this conference was convened to honor, was to apply the Einstein-Infeld-Hoffmann approach to deriving the equations of motion for the wave/particle as a corollary to the field equation for the spacetime Lorentzian metric tensor field. (cf., [**17**].)

Since the the linear wavelike nature of the electromagnetic field was derivable from the linearity of Maxwell's equations, one should examine not only nonlinear wave mechanics as possible extension of traditional quantum mechanics, but also nonlinear electrodynamics [**18**].

One of the early attempts at such a theory was made by Gustav Mie [**19**]. What he achieved was more of a program for nonlinear electrodynamics, since he mostly elaborated upon the set of Lorentz invariant expressions that one could form from the Minkowski field strength 2-form $F$, its Hodge dual $*F$, and the potential 1-form $A$. Since one could then use these expressions in the construction of electromagnetic field Lagrangians, he then explored the properties of some possible Lagrangians. It is interesting that some of the early objections to his work seem less convincing in the light of modern QED. For instance, the inclusion of terms in $A$, which would break the $U(1)$ gauge invariance of the Lagrangian, is closely related to the appearance of a possible non-zero photon mass. Futhermore, one of his models, which seemed physically absurd at the time, describes a process that now suspiciously resembles pair creation and annihilation.

Perhaps the most definitive theory of nonlinear electrodynamics to date was the one that was conceived by Born and Infeld [**20**]. The essential physical element that it included was a critical field electric field strength $E_c$, beyond which the phase transition of vacuum polarization would occur. Another, subtler, aspect of the Born-Infeld Lagrangian:

$$\mathscr{L} = \sqrt{E_c^2 + \mathscr{F}^2 - \mathscr{G}^2} - E_c, \qquad (3.2)$$

was the fact that in addition to the kinetic energy term $\mathscr{F}^2 = F\wedge *F$ that one expects in a gauge field theory, the Lagrangian also included the topological charge density $\mathscr{G}^2 = F\wedge F$. One reason for the continued interest in the Born-Infeld Lagrangian, even in this age of quantum field theory, is that it very closely represents an effective Lagrangian for the vacuum polarization behavior that was derived by Euler and Heisenberg [**21,22**] by starting with the Dirac equation.

If vacuum polarization actually changes the optical properties of the spacetime vacuum manifold – i.e., $\varepsilon_0$ and $\mu_0$ – in the realm of large electric field strengths in such a way that the resulting electromagnetic wave motion is nonlinear then perhaps one possible phenomenological bridge between quantum electrodynamics and nonlinear electrodynamics is by way of nonlinear optical analogies. For instance, many of the elementary scattering processes that are treated in quantum electrodynamics have analogs in nonlinear optics, such as the scattering of light by light or external potentials. However, in order to make the analogies physically meaningful, one must also address the properties of the spacetime electromagnetic vacuum manifold as an optical medium in deeper detail.

**4. Spacetime as an ordered medium.** A consistent picture that seems to have been emerging throughout the foregoing discussion is the possibility of treating spacetime as an ordered medium that exists in various phases and is capable of exhibiting phase transitions, as well.

The branch of mathematics that seems to say the most about the concept of order is group theory, and, in particular, the way that symmetries arise under the action of transformation groups. In condensed matter [**23-30**], a consistent vocabulary for the structure of ordered media seems to be gradually settling out. It is based in the common elements that one observes in the phase transitions of ferromagnetism, superconductivity, superfluidity, liquid crystal states, and the onset of turbulence in viscous fluids, to name a few such phenomena. Various attempts have been made to apply these models to the more esoteric problems of elementary particle physics and cosmology, such as the Nambu-Jona-Lasinio hypothesis that mesons might be analogous to the quasiparticle pairs in the BCS theory of superconductivity, and the way that the early years of string theory were partly based in Nambu's suggestion that the Nielsen-Olesen vortices of two-dimensional gauge field theory might be analogous to the Abrikosov vortices in type-II superconductors.

A crucial element of the structure of ordered media is the fact that its phase transitions are associated with a "spontaneous" reduction of the symmetry group of the ground state to a subgroup without affecting the symmetry of the overall Lagrangian (or free energy) of the system. For instance, when a ferromagnetic medium goes from a non-magnetized phase into a magnetized one, the ground state goes from the *SO*(3) rotational symmetry of an isotropic medium to the *SO*(2) symmetry of a medium in which a preferred direction exists. The order parameter for this medium is then the unit vector field in space that defines the direction of magnetization at each point. It takes its values in the unit sphere of angular direction coordinates, which also happens to be the homogeneous space *SO*(3)/*SO*(2). More generally, when a phase of a medium that occupies a region of space described by a differentiable manifold *M* is defined by the reduction from a Lie group *G* to a subgroup *H*, the order parameter for the medium is a smooth function from *M* to *G/H*. The homogeneous space *G/H* is often called the *vacuum manifold* for that phase. One further refines this picture of ordered media by describing the homotopy classes [*M*; *G/H*] as the *topologically stable* ground state configurations for that phase. When one uses the Lie groups *G* and *H* as the gauge groups of internal symmetries in gauge field theories, one also finds that the process of spontaneous symmetry breaking is associated with the reduction of the *G*-principal bundle that defines the *G*-gauge structure for the field to an *H*-principal bundle.

In the case of the spacetime manifold, when one considers it to be a four-dimensional differentiable manifold *M* the symmetry groups that appear most naturally are the subgroups of the affine group for four dimensions, *A*(4), and the *G*-principal bundles that they describe are usually reductions of the bundle of affine frames, *A*(*M*), to *G*-principal sub-bundles. In particular, they are generally reductions of the bundle of linear frames *GL*(*M*).

In the general case, if *M* is an *n*-dimensional differentiable manifold then a reduction of *GL*(*M*) to a *G*-principal bundle, where *G* is a subgroup of *GL*(*n*) is called a *G-structure* [**31-33**]. The elements of a *G*-structure at each point $x \in M$ will then be the linear frames in $T_x(M)$ that all lie within the same orbit under the (right) action of *G* on frames. The examples of *G*-structures seem to encompass all that is holy in differential geometry. For instance, when $G = SL(n)$, the frames at each point will all have the same (by definition, unit) volume (i.e., determinant) and an *SL*(*n*)-structure on *M* is associated with a global volume element. This volume element gives us an example of a *G*-structure that is defined by a *fundamental tensor field*, which can be described as a *G*-equivariant map *t*: $GL(M) \to GL(n)/G$ that is constant on *G*(*M*).

The latter example already points to two key issues in the process reducing principal bundles: their existence and their uniqueness, at least up to some form of equivalence, such as homotopy. In the present case, in order for *GL*(*M*) to admit a reduction to an *SL*(*n*)-principal bundle, it must be orientable. If *M* is compact then this is equivalent to the vanishing of the first Stiefel-Whitney class of its tangent bundle, $w_1[M] \in H^1(M; \mathbb{Z}_2)$. In order for this to be true, it is sufficient, but not necessary, that *M* be simply connected. However, since the homotopy classes of orientations will be indexed by $H^0(M; \mathbb{Z}_2)$, as long as *M* is path-connected, there will be only two orientations, up to homotopy.

Other examples of *G*-structures that may or may not exist without topological obstructions are Riemannian structures ($G = O(n)$, fundamental tensor field = *g*, no obstructions), Lorentzian structures ($G = O(n\text{-}1, 1)$, fundamental tensor field = *g*, obstruction = Euler-Poincaré characteristic for a compact *M*, no obstruction for non-compact *M*), and ultimately a parallelization of *M*, i.e., a global frame field $\mathbf{e}_\mu$ ($G = \{e\}$, fundamental tensor field = $\mathbf{e}_\mu$, topological obstruction = vanishing of all Stiefel-Whitney classes). One can even regard symplectic structures, differential systems, and almost-complex structures as *G*-structures.

Along the aforementioned topological issues, one also can consider geometrical ones. In particular, suppose one starts with a linear connection 1-form ω on *GL*(*M*), whose torsion and curvature satisfy the Cartan structural equations:

$$\begin{cases} \Theta^\mu = d\theta^\mu + \omega^\mu_\nu \wedge \theta^\mu \\ \Omega^\mu_\nu = d\omega^\mu_\nu + \omega^\mu_\lambda \wedge \omega^\lambda_\nu, \end{cases} \quad (4.1)$$

in which $\theta^\mu$ is the canonical $\mathbb{R}^\nu$-valued 1-form on $GL(M)$ (for a local frame field $\mathbf{e}_\mu$ on $U \subset M$ it pulls down to the reciprocal coframe field to $\mathbf{e}_\mu$, which is what one usually considers in the classical relativity literature). If one reduces $GL(M)$ to a $G$-structure $G(M)$ then the main geometrical issue is whether all of the geometrical information, namely, $\theta^\mu$, $\omega$, $\Theta$, $\Omega$, can be reduced to corresponding information on $G(M)$. In order to reduce $\theta^\mu$ all one must do is restrict its definition to the $G$-frames, so this reduction is automatic. However, the situation is more involved with $\omega$, since $\omega$ must take its values in the Lie algebra of the structure group of the principal bundle that it is defined on. Hence, the only way that a $GL(n)$-connection can reduce to a $G$-connection when $G$ is a proper subgroup of $GL(n)$ is for $\omega$ to take its values completely within the Lie algebra $\mathfrak{g}$ to begin with. When $G(M)$ is associated with a fundamental tensor field $t$ a linear connection $\omega$ is reducible to a $G$-connection iff:

$$\nabla t = dt + \omega \wedge t = 0, \tag{4.2}$$

in which we are leaving aside any specification of the way that the Lie algebra $\mathfrak{g}$ acts on the fundamental tensor field and abbreviating that action by $\omega \wedge t$.

When $G = O(n)$ or $O(n-1, 1)$, so $t$ is really the metric tensor field $g$, then this latter condition for the reducibility of a linear connection takes the familiar form of demanding that the metricity tensor $Q = \nabla g$ must vanish. This also gives us a convenient example for examining the contrary case, since one also has that if a linear connection $\omega$ is not reducible to a metric connection then one can express it in the form:

$$\omega = \varpi + \tau, \tag{4.3}$$

in which $\varpi$ is a metric connection and $\tau$ is a 1-form with values in a linear subspace of $\mathfrak{gl}(n)$ that is complementary to $\mathfrak{o}(n)$ or $\mathfrak{o}(n-1,1)$, respectively.

If $\omega$ is reducible then the torsion and curvature of its reduction are obtained by restriction; otherwise, they too will be decomposable in a manner that is analogous to (4.3). It is convenient to think that when a $G$-structure is defined by a fundamental tensor field $t$, the basic geometric data must be augmented to $(t, \theta^\mu, \varpi)$ and the associated structure equations then get augmented to equations for $(Q, \Theta, \Omega)$, respectively. More generally, when $\varpi$ is obtained from an irreducible connection $\omega$ the structure equations for $\varpi$ must be supplemented with *equations of deformation* for $t$.

Now, let us bring this esoteric mathematical discussion back to the case at hand of how one can define an analogy between the ordered media of condensed matter physics and the $G$-structures that may or may not exist globally on the spacetime manifold $M$ ($n = 4$). The approach that the author has been taking [**34**] is to regard the subgroup $G$ in $GL(4)$ as representing a *phase* of the spacetime manifold that is defined by the way that $G$ acts as an internal symmetry group on any $G$-structure $G(M)$, and a reduction from $G$ to a subgroup $H$ represents a *phase transition* that is associated with the spontaneous breaking of the symmetry $G$ to $H$. The spacetime *vacuum manifold* for the $H$ phase is the homogeneous space *G/H* that is associated with the reduction, and the *order parameter* for that phase is the fundamental tensor field that is associated with the reduction. The topological obstructions to the reduction seem to play a role that is analogous to, but somewhat more general than, the topological defects in ordered media. The role of the 1-form τ that arises when the *G*-connection ω does not reduce to an *H*-connection seems to be related to the deformations of the geometry of an ordered medium that surround topological defects. Since the curvature of a connection is also an obstruction to the integrability of the differential system that is defined by the horizontal sub-bundle of the tangent bundle to *G*(*M*), it is conceivable that a closer study of the integrability of *G*-structures will uncover a link between the topological and geometrical nature of topological defects that might define the defects as the sources of the deformations.

Naturally, a long-range goal of the aforementioned program is to see if the problems of gravitation, electromagnetism, and wave mechanics can be posed in the consistent language of *G*-structures in such a way that one might unify all three theories. In the pre-metric formulation of electromagnetism [**35,36**], the appropriate setting seems to be an *SL*(4)-structure on spacetime − i.e., an orientation and global volume element – which allows one to define Poincaré duality, but not Hodge duality. In order to define Hodge duality one also needs a metric tensor field, i.e., a reduction of *SL*(*M*) to an *SO*(3,1)-structure. However, in order to define this reduction by starting in electromagnetism, it turns out that it is sufficient to define a linear constitutive law, which amounts to a metric χ on the bundle of 2-forms $\Lambda^2(M)$. This probably needs to be extended to nonlinear constitutive laws, but such a structure is closely related to defining a scalar product on the second de Rham cohomology $H^2(M; \mathbb{R})$, since every closed 2-form is associated with a cohomology class in $H^2(M; \mathbb{R})$. A topological essential example of a scalar product on $H^2(M; \mathbb{R})$ is given the *intersection form* of *M* [**37-39**], which seems to embody all of the other topological information, at least when *M* is simply connected ([3]).

Of course, the reduction of *SL*(*M*) to a bundle of oriented Lorentzian frames also brings one into the arena of gravitation, which centers on the Lorentzian metric tensor field *g* that also defines the fundamental tensor field of the reduction. Hence, one already has a considerable inventory of geometrical insights concerning that phase of the spacetime vacuum manifold.

Apparently, an essential aspect of wave motion in a four-dimensional manifold is the fact that it is associated with a reduction of *GL*(*M*) to an *SO*(2)-structure [**40**]. This is because in order to describe wave motion one must generally define two orthogonal unit vector fields **t** and **n**, which define the direction of proper time evolution and the normal vectors to the isophase hypersurfaces, respectively. This fixes a two-dimensional sub-bundle of *T*(*M*), as well as its orthogonal complement Φ(*M*), and defines a class of Lorentz-orthonormal frames that all have **t** and **n** as members. That leaves the other two spacelike orthonormal vectors, which live in Φ(*M*), defined only up to a phase angle. Hence, we have reduced the bundle of oriented Lorentzian frames to a bundle of oriented spacelike orthonormal 2-frames whose structure group is *SO*(2).

---

[3] In the contrary case, one would define such an object on the simply-connected orientable covering manifold to *M* and examine the physical significance of projective singularities.

Ultimately, one must ask what role the final reduction to a global – or at least partial – frame field on spacetime plays. Traditionally, this is the setting for *teleparallelism*, which was once considered to be a possible unification of gravitation and electromagnetism. It is also the setting for the translational gauge theories of gravitation. It is interesting that none of the physical literature on teleparallelism seems to address the topological fact that parallelizability is actually rather hard to come by and involves a high degree of symmetry in the manifold; indeed, one cannot even define a global frame field on such a symmetric and homogeneous manifold as $S^2$. Note that one of the foundational papers on the subject of parallelizability (cf. Stiefel [**41**]) was not published until several years after Einstein and Mayer abandoned teleparallelism as a means of unifying the theories of gravitation and electromagnetism. Of course, whether this decision was premature because more subtle topological considerations might have resolved the conflict that led them to give up is yet to be determined.

## Acknowledgements

The author wishes to thank Jean-Pierre Vigier for the inspiration the author derived from reading Vigier's book on the causal interpretation of wave mechanics, Friedrich Hehl at the University of Cologne for directing the author's attention to the pre-metric formulation of electromagnetism, and the University of Wisconsin at River Falls for its generous support in allowing the author to attend the Vigier IV conference.